\documentclass[%
preprint,
 amsmath,amssymb,
 aps,
]{revtex4-1}

\usepackage{graphicx}
\usepackage{dcolumn}
\usepackage{bm}

\begin{document}

\preprint{}
\title{Thin liquid film resulting from a distributed source \\ on a vertical wall}

\author{Yadong Ruan}
 \email{yadong.ruan@cgu.edu}
\affiliation{%
 Institute of Mathematical Sciences\\
 Claremont Graduate University\\
 Claremont, CA 91711 USA
}%

\author{Ali Nadim}
 \email{ali.nadim@cgu.edu}
\affiliation{%
 Institute of Mathematical Sciences\\
 Claremont Graduate University\\
 Claremont, CA 91711 USA
}%

\author{Marina Chugunova}
 \email{marina.chugunova@cgu.edu}
\affiliation{%
 Institute of Mathematical Sciences\\
 Claremont Graduate University\\
 Claremont, CA 91711 USA
}%




\date{\today}

\begin{abstract}
We examine the dynamics of a thin film formed by a distributed liquid source on a vertical solid wall. The model is derived using the lubrication approximation and includes the effects of gravity, upward airflow and surface tension. When surface tension is neglected, a critical source strength is found below which the film flows entirely upward due to the airflow, and above which some of the flow is carried downward by gravity. In both cases, a steady state is established over the region where the finite source is located. Shock waves that propagate in both directions away from the source region are analysed. Numerical simulations are included to validate the analytical results. For models including surface tension, numerical simulations are carried out. The presence of surface tension, even when small, causes a dramatic change in the film profiles and the speed and structure of the shock waves. These are studied in more detail by examining the traveling wave solutions away from the source region.
\end{abstract}

\maketitle

\onecolumngrid
\section{Introduction} \label{sec:intro}
The motivation for this work was an industrial problem presented by W.~L. Gore and Associates at the Mathematical Problems in Industry (MPI) workshop that took place in Claremont, CA in June of 2018. The problem concerned modeling dense porous catalysts in which a gaseous reaction produces liquid in the interior of the catalyst, which gradually pushes its way out to the exterior surface, forming drops or films of liquid on that surface. These block the gaseous reactants from entering the pores and slow down the reaction. In order to remove the liquid drops or films from the surface, one option being considered was to temporarily increase the flow of gas past the surface in the hope of blowing off the liquid film. 

In this work, in order to gain insight into some of the underlying physics of that problem, we undertook to model a thin liquid film on a vertical wall, being generated by a finite distributed source of liquid on the wall to represent the liquid that oozes out of the porous catalyst onto the surface. We included the effects of gravity which causes the film to flow downward along the wall, as well as an upward airflow that, if strong enough, could drag the film up the wall. We also included the effects of surface tension in our model.

The evolution of film thickness driven by various external driving forces is of much interest given its applications in many different areas of physics and engineering involving coating flows. In such flows, if the film is thin in one dimension compared to the others, the so-called lubrication approximation provides a simpler model for analysis, as opposed to solving the full Navier-Stokes equations that govern viscous fluid flow. A review of lubrication theory is provided in \cite{o2002theory}. Models in higher dimensions are also being investigated, such as the three-dimensional gravity-driven thin liquid film flow on an inclined plane described in \cite{lan2010developing}.

In some of the thin film mathematical models, solutions of particular form can be constructed, including travelling wave, similarity and steady state solutions. These solutions provide insights for further analysis. For instance, in \cite{duffy1997similarity}, the authors provided a similarity solution for viscous source flow on inclined plane. Certain properties of the derived thin film model, such as the speed of drop spreading, are also important. For example, analyse of the minimum wetting rate and the corresponding minimum liquid film thickness were presented in \cite{el2001minimum} and validated with experimental data. 

Another aspect of these problems that has received a lot of attention is the stability of the film under different perturbations as well as methods to stabilize the film. Stability of thin wavy films flowing down an inclined plate was studied in \cite{nepomnyashchii1974stability} and \cite{krantz1971stability}. In \cite{davalos2007nonlinear} and \cite{davalos2008instabilities}, the author introduced several functions for deformed walls to stabilize the film surface with respect to time-dependent perturbations, reporting numerical results. The stability of liquid flow down a heated inclined plane was examined in \cite{bankoff1971stability}. References \cite{cheng2000stability} and \cite{cheng2001nonlinear} respectively studied thin viscoelastic liquid films flowing down a vertical wall and a vertical cylinder. In \cite{sadiq2008thin}, the stability of liquid film flow on a porous inclined plane was examined, while the film stability on a wavy surface was studied in \cite{wierschem2003instability}. 

Many experiments and theoretical analyses have been done on the motion of thin films with a given initial condition. For instance, an accelerating laminar thin-film flow along a vertical wall was investigated in \cite{roy1984laminar}, laminar flow on a wavy inclined surface was studied in \cite{bontozoglou1997laminar}, and liquid films falling vertically on the outer wall of a circular tube were studied experimentally in \cite{takahama1980longitudinal}. Several characteristics of thin film flow on inclined surfaces were studied in \cite{brauner1982characteristics}. Three-dimensional droplet models and wave dynamics on inclined and vertical walls were studied in \cite{schwartz2005shapes} and \cite{park2003three}, respectively. Experimental studies of viscous, particle-laden thin films were reported in \cite{ward2009experimental}. Flows under obstacles were examined in \cite{blyth2006film}. 

Few authors have considered source terms in the thin film equation. In \cite{higuera1995steady}, a numerical method for the Reynolds equation for a steady liquid layer flowing down a slightly inclined plate from a point source is presented. In \cite{lister1992viscous}, the flow of a viscous fluid from a point or line source on an inclined plane is analyzed. The effect of surface tension was neglected in \cite{lister1992viscous}. In our present work, we model thin films formed from a finite source region along a vertical solid wall while considering the effects of gravity, airflow and surface tension. This case is important since some industrial gaseous chemical reactions that occur in porous catalysts give rise to liquids on the exterior surfaces that fit within this model. 

Some research has also covered non-Newtonian fluids, including thin-film flow of a power-law liquid on an inclined plate in \cite{miladinova2004thin} and stability analysis of travelling wave solutions of power-law liquid films in \cite{perazzo2004steady}. A exact solution of the thin film flow problem for a third grade fluid on an inclined plane is provided in \cite{hayat2008exact}. 

In this paper,  we derive a mathematical model for film motion along a solid vertical wall in the form of a partial differential equation for film thickness $h(x,t)$ as a function of distance $x$ increasing downward and time $t$. The final model after scaling turns out to be 
\begin{align}
    h_t + ( h^3 - h^2 + \alpha h_{xxx})_x = S(x)\,, \label{ full model}
\end{align}
where $\alpha > 0$ is a dimensionless parameter that characterizes the effect of surface tension. Terms $h^3$ and $-h^2$ represent the downward flux due to gravity and the upward flux due to airflow, respectively. The right-hand side in this equation is the distributed source which we take to be of the form 
\begin{equation}
	S = \begin{cases} S_0 & \text{ if } x \in (0,1)\\
	0 & \text{ otherwise. }
	\end{cases}
	\label{source}
\end{equation}

The main results of this paper are in two parts. In the first part we consider the case $\alpha = 0$, which is the case where surface tension can be ignored. In many practical conditions, the dimensionless surface tension parameter is indeed very small. The model then reduce to 
\begin{align}
    h_t + ( h^3 - h^2)_x = S(x)\,.
\end{align}
For this first order nonlinear partial differential equation, we use the method of characteristics to analyze its dynamics. With the source given by Eq.~($\ref{source}$), shock waves will form, with their number and structure depending on $S_0$. For all values of $S_0$, an upward propagating shock wave will form as the film is carried up by the airflow. However, we obtain a critical source value $S_c$, so that if $S_0 > S_c$, a second downward propagating shock wave will also form, as the excess fluid falls downward due to gravity. A steady state solution over the source region $(0, 1)$ is also derived, with dramatically different form depending on whether the source strength is below or above the threshold value. A numerical solution is also obtained to validate the results from the method of characteristics. 

In the second part of this work, we consider the full model with surface tension effects. Numerical simulations are carried out for various $S_0$ and $\alpha$ values. The numerical results indicate potential connections between the solutions of the full model (\ref{ full model}) with travelling wave solutions of the thin film equation without source. Importantly, we find that even for quite small values of the surface tension parameter $\alpha$, there is a significant change in the profile of the thin film and the speed of the shock waves, as compared to the case with zero surface tension.

\section{Model Derivation} \label{sec:derive}
We model a thin-film driven by gravity and external airflow under the lubrication approximation. We assume the flow to be two-dimensional with coordinate $x$ along the wall and $y$ normal to the wall, with respective velocity components $u$ and $v$, and take the wall to make angle $\alpha$ with the horizontal direction, which for a vertical wall will become $\alpha=\pi/2$. Let us start with the Navier-Stokes equations with constant viscosity $\mu$ and density $\rho$:
\begin{align}
	\rho ( \frac{\partial \vec{u}}{\partial t} + \vec{u} \cdot \nabla \vec{u} ) &= \rho \vec{g} - \nabla p + \mu \Delta \vec{u} \,. \label{eq:1}
\end{align}
Denote the scale of fluid velocity components $\vec{u} = (u,v)^T$ by $U$ and $V$ respectively, the scale of film thickness by $H$, and that of the $x$ domain by $L$. To apply lubrication approximation, we need
\begin{align}
	\epsilon = \frac{H}{L} \ll 1\,.
\end{align}
The continuity equation for an incompressible liquid reads  
\begin{align}
	\nabla \cdot \vec{u} = 0 \quad \Rightarrow \quad \frac{\partial u}{ \partial x} + \frac{\partial v}{\partial y} = 0\,.
\end{align}
Since the continuity equation needs to be satisfied exactly, upon balancing the respective scales of the two terms we find 
\begin{align}
	\frac{U}{L} = \frac{V}{H} \quad \Rightarrow \quad V = \frac{H}{L}U = \epsilon U\,. \label{eq:uv}
\end{align}
Now from the $x$-component of equation (\ref{eq:1}): 
\begin{align}
	\rho \frac{\partial u}{\partial t} + \rho u \frac{\partial u}{\partial x} + \rho v \frac{\partial u}{\partial y} = \rho g \sin{\alpha} - \frac{\partial p }{\partial x} + \mu (\frac{\partial^2 }{\partial x^2} + \frac{\partial^2 }{\partial y^2}) u \,,
	\label{eq:2}
\end{align}
and using $T$ as the scale for time $t$ and $P$ as that for pressure $p$, the scales of the seven terms in that equation, in order, become
\begin{align}
	\frac{\rho U}{T}\,, \quad \frac{\rho U^2}{L}\,, \quad \frac{\rho V U}{H}\,, \quad  \rho g\,, 
	\quad \frac{P}{L}\,, \quad \frac{\mu U}{L^2}\,, \quad \frac{\mu U}{H^2}\,. 
\end{align}
Since ${H}/{L} \ll 1$, the last term on the RHS of (\ref{eq:2}) is dominant with scale $\mu{U}/{H^2}$, and the term ${\mu U}/{L^2}$ is smaller by a factor of $\epsilon^2$. To keep the pressure term in balance with the dominant term, we need the scale $P$ for pressure to be 
\begin{align}
	P = \frac{\mu UL}{H^2} = \frac{\mu U}{ \epsilon^2 L}\,.
\end{align}
Also for the gravity term to be of similar magnitude: 
\begin{align}
	\rho g \sim \frac{\mu U}{H^2} \quad \Rightarrow \quad U = \frac{\rho g H^2}{\mu}
\end{align}
which determines the scale $U$ of velocity in the $x$ direction under the model that includes gravity.

On the LHS of (\ref{eq:2}), the second and third terms have scales ${\rho U^2}/{L}$ by using the result from (\ref{eq:uv}). We choose the time scales $T$ as 
\begin{align} 
	T = \frac{L}{U}
\end{align}
which is the characteristic time for the flow to traverse a distance $L$ at speed $U$. As such, all the LHS terms have scale ${\rho U^2}/{L}$ and the ratio of the LHS to RHS scales turns out to be 
\begin{align}
	\frac{ {\rho U^2}/{L} }{ {\mu U}/{H^2} } = (\frac{H}{L})^2 \frac{\rho U L}{\mu} = \epsilon^2 \text{Re}_L \,,
\end{align}
where $\text{Re}_L={\rho U L}/{\mu}$ is the Reynolds number. Under the assumption that $\epsilon^2 \text{Re}_L \ll 1$, the inertia terms on the LHS of the momentum equation are negligible compared to the terms on the RHS. Hence, to leading order, we can approximate the $x$-momentum equation by: 
\begin{align}
\rho g \sin{\alpha} - \frac{\partial p }{\partial x} + \mu  \frac{\partial^2 u}{\partial y^2} = 0\,.
\end{align}
Similarly, the $y$-component of equation (\ref{eq:1}) with the same scaling applied to all the terms results in the leading order equation:
\begin{align}
	 0 = -\rho g \cos \alpha - \frac{\partial p}{\partial y}\,.
\end{align}

We now discuss the boundary conditions on the solid-liquid ($y = 0$) and liquid-air ($y = h(x, t) $) interfaces. No-slip and no-penetration conditions at the solid-liquid interface would normally require: 
\begin{align}
	u(x, 0, t) = 0, \quad v(x, 0, t) = 0\,.
\end{align}
However, when a steady fluid source is considered at the interface, with liquid volume emanating from the porous wall, the condition on the velocity component $v$ changes to $v(x, 0, t) = S(x)$ where $S(x)$ is the source strength. 

At the liquid-air interface $y = h(x, t) $, we have kinematic and dynamic boundary conditions. The normal stress balance at the interface reads:
\begin{align}
	\hat{n} \cdot [\pi_{\tiny \mbox{air}} - \pi_{\tiny \mbox{liquid}}] \cdot \hat{n} = \sigma \mathcal{K} \label{eq:normal-stress-balance}
\end{align}
where $\hat{n}$ is the normal vector pointing from the liquid towards the air, and $\mathcal{K}$ is the local curvature of interface, and we have 
\[
	\pi_{\tiny\mbox{air}} =  -p_{\tiny\mbox{atm}}I, \quad \pi_{\tiny\mbox{liquid}} =  -p I + \mu \begin{pmatrix}
	2u_x & v_x + u_y \\
	v_x + u_y & 2v_y
	\end{pmatrix}
\]
The normal vector is well approximated by the unit vector in the $y$-direction since ${\partial h}/{\partial x}$ has scale ${H}/{L} = \epsilon \ll 1$:
\begin{align}
	\hat{n} = \frac{ \nabla ( y - h(x,t))}{|\nabla (y - h(x,t))|} = \frac{1}{\sqrt{1 + (\frac{\partial h}{ \partial x})^2}} \begin{pmatrix}
- \frac{\partial h}{\partial x} \\
1
	\end{pmatrix}  \approx \begin{pmatrix}
0 \\
1
	\end{pmatrix} 
\end{align}
For curvature $\mathcal{K}$ we have
\begin{align}
	\mathcal{K} = \nabla \cdot \hat{n} \approx -\frac{\partial^2 h}{ \partial x^2}
\end{align}
Substituting into (\ref{eq:normal-stress-balance}), we find 
\begin{align}
	-p_{\tiny\mbox{atm}} + p  - 2\mu \frac{\partial v}{\partial y} = - \sigma \frac{\partial^2 h}{\partial x^2}\,.
\end{align}
In order for the surface tension term not to be negligible, we need the scale for last term to be comparable to pressure terms; that is, we must have
\begin{align}
	\frac{\sigma H}{ L^2} \sim \frac{1}{\epsilon^2} \frac{\mu U}{L} \quad\Rightarrow\quad \frac{\mu U}{\sigma} \sim \epsilon^3 \,.
	\label{surface_tension_condition}
\end{align}
This corresponds to having a very small capillary number, requiring surface tension to be relatively large compare to viscous effects. Under this scaling and recognizing that the normal viscous stress $\mu (\partial v/\partial y)$ is also small compared to the other terms, the normal stress balance simplifies to  
\begin{align}
	p - p_{\tiny\mbox{atm}} = - \sigma \frac{\partial^2 h}{\partial x^2}\,.
\end{align}
Now consider the tangential stress balance at the interface which reads
\begin{align}
	\hat{n} \cdot \pi_{\tiny\mbox{liquid}}\cdot \hat{t}  + \tau =0 \label{eq:tangent-stress-balance}
\end{align}
where $\tau$ is the upward wind stress exerted by the external airflow and  $\hat{t} \approx (1,0)^T$ is the unit tangent at the interface. This equation reduces to
\begin{align}
	\mu( \frac{\partial u}{ \partial y} + \frac{\partial v}{\partial x}  ) = -\tau\,,
\end{align}
which, given that the scale of ${\partial u}/{ \partial y}$ is much larger than that of ${\partial v}/{\partial x}$, simplifies to 
\begin{align}
\mu \frac{\partial u}{ \partial y} = -\tau\,.
\end{align}
The kinematic boundary condition at the interface requires
\begin{align}
	\frac{D}{Dt}(y - h(x,t)) = 0 \quad\Rightarrow\quad (y - h(x,t))_t + \vec{u} \cdot  \nabla ( y- h(x,t)) = 0\,.
\end{align}
This results in
\begin{align}
	\frac{\partial h}{\partial t} = v - u \frac{\partial h}{\partial x} \,.\label{kinematic-BC}
\end{align}
Based on the scales we determined earlier, including the one for time $t$, we see that all three terms have comparable scales $\epsilon U$.

Summarizing all the equations and boundary conditions and specializing to the case when the wall is vertical, i.e., $\alpha=\pi/2$, we have:
\begin{align}
	0 &= \frac{\partial p}{\partial y} \label{NS-y}\\
	0 &= \rho g  - \frac{\partial p}{\partial x} + \mu \frac{\partial^2 u}{\partial y^2} \label{NS-x}\\
	0 &= \frac{\partial u}{\partial x} + \frac{\partial v}{\partial y}  \label{continuity}
\end{align}
which respectively represent the $y$- and $x$-components of the momentum equation and the continuity (incompressibility) equation, subject to boundary conditions at $y = 0$: 
\begin{align}
	u &= 0  \label{no-slip-BC}\\
	v &= S(x)  \label{no-penetration-BC}
\end{align}
and those at $y = h(x,t)$: 
\begin{align}
	p &= p_{\tiny\mbox{atm}} - \sigma \frac{\partial^2 h}{\partial x^2}   \label{normal-BC}\\
	\mu \frac{\partial u}{\partial y} &= -\tau\,.  \label{tangent-BC}
\end{align}

Differentiating (\ref{normal-BC}) with respect to $x$, we find
\begin{align}
	\frac{\partial p}{\partial x} = - \sigma \frac{\partial^3 h }{\partial x^3}\,.
\end{align}
 This term is also independent of $y$ because of equation (\ref{NS-y}). Integrating (\ref{NS-x}) with respect to $y$ twice, we obtain
\begin{align}
	u(x,y,t) = \frac{1}{\mu}(-\sigma \frac{\partial^3 h}{\partial x^3} - \rho g) \frac{y^2}{2} + \frac{1}{\mu}C_1(x,t)y + C_2(x,t)\,.
\end{align}
Using (\ref{no-slip-BC}), we have $C_2(x,t) = 0$ and using (\ref{tangent-BC}), we find 
\[
	C_1 = \frac{h}{\mu}( \rho g + \sigma \frac{\partial^3 h}{\partial x^3}) - \frac{\tau}{\mu}\,.
\]
Integrating (\ref{continuity}) at a fixed location $x$ with respect to $y$ from 0 to $h(x,t)$, and making use of (\ref{no-penetration-BC}) and (\ref{kinematic-BC})  yields:
\begin{align}
	\frac{\partial h}{\partial t} + \frac{\partial q}{\partial x}  = S(x)\,,
\end{align}
where the volume flux $q$ has been defined as $q = \int_0^h u(x,y,t) dy$. The latter can be found from the velocity profile given above to have the explicit form:
\begin{align}
q &= \frac{\rho g}{3\mu} h^3 - \frac{\tau}{2\mu} h^2 + \frac{\sigma}{3\mu} \frac{\partial^3 h}{\partial x^3} h^3\,.
\end{align}
The first term on the RHS represents the downward flow due to gravity and the second term the updard flow due to the airflow. If surface tension is not as large in magnitude as required by the scaling (\ref{surface_tension_condition}), we can ignore the effects of surface tension and drop the last term in the expression for the flux. 

While we derived the above conservation equation and flux expression in dimensional form, albeit guided by the scaling analysis which indicated which terms could be neglected, at this point we can go ahead and nondimensionalize the system. Define the starred dimensionless variables by
\begin{align}
	h = H h^* , \quad x = Lx^*, \quad t = T t^*  , \quad S = S_{\tiny\mbox{scale}}S^* 
\end{align}
with 
\begin{align}
	H = \frac{3 \tau}{2 \rho g}, \quad T = \frac{4 \mu \rho g L}{3 \tau^2}, \quad S_{\tiny\mbox{scale}} = \frac{9 \tau^3}{ 8 \mu \rho^2 g^2 L}\,.
\end{align}
Here, length scale $H$ corresponds to the film thickness at which the downward flux due to gravity exactly balances the upward flux due to the wind stress associated with airflow; i.e., the film thickness at which the first two terms in the expression for flux $q$ balance each other exactly. The length scale $L$ is associated with the distance along the wall, for instance the length of the region over which the source is nonzero. By assumption, $\epsilon=H/L \ll 1$. The time scale $T$ in the above can be shown to be equivalent to $T=3L/U$ with velocity scale $U$ given by $U=\rho g H^2/\mu$. The scale for the source emerges naturally from equating the orders of magnitude of the terms in the conservation equation. Substituting these and and dropping the superscript star from the dimensionless variables for clarity, we finally have
\begin{align} \label{PDE_with_alpha}
	h_t + (h^3 - h^2 + \alpha \, h^3 h_{xxx})_x = S(x)
\end{align}
where $\alpha=\sigma H /(\rho g L^3)=\epsilon^3/\mbox{Ca}$, where $\mbox{Ca}=\mu U/\sigma$ is the capillary number based on the velocity scale $U=\rho g H^2/\mu$. As indicated earlier, in order for surface tension not to be negligible, the Capillary number needs to be small, of order $\epsilon^3$, which would make dimensionless parameter $\alpha$ of order unity. The model we derived here is similar to the thin film model with gravity and Marangoni effects in \cite{PhysRevE.96.043107} and \cite{Bertozzi1999}. In the next section, we first analyze the case where surface tension effects are negligible, by taking coefficient $\alpha$ to be zero. However, since that coefficient multiplies the highest order spatial derivative, one can expect a somewhat singular behavior so that the solution in the presence of $\alpha$, no matter how small, might be qualitatively different from that in the complete absence of surface tension. We will see that this is indeed the case in a later section where surface tension effects are added back in.

\section{Model without Surface Tension}
To ignore the effect of surface tension, parameter $\alpha$ is set to zero. Furthermore, the source strength $S(x)$ is assumed to be uniform over a finite domain of dimensionless length 1, and zero elsewhere, namely: 
\[
	S(x) = \begin{cases} S_0 & \text{ if } x \in [0,1] \\
	0 & \text{ otherwise }
	\end{cases}
\]
In this case, we can derive certain results through analysis. We will find that if the source strength $S_0$ is less than a threshold, the liquid is carried upward by the airflow and none of it falls down due to gravity. The upper front of the film propagates as a shock front, whose speed we can predict. When the source strength exceeds the threshold, some of the liquid produced is still carried upward by the airflow, while the rest falls down due to gravity. Over the region where the source is nonzero, a steady film profile is achieved in both cases. A numerical solution of the nonlinear film equation produces results that agree with the analytical predictions. 
\subsection{The Simplified Model}
In the absence of surface tension, the expression for the flux becomes $q(h)=h^3-h^2$ and the film thickness $h(x, t)$ satisfies the simplified equation: 
\[
\frac{\partial h}{ \partial t} + (3h^2-2h)\frac{\partial h}{ \partial x} = S(x)\,,
\]
with initial condition 
\[
	h(x, 0) = 0\,,
\]
corresponding to not having any liquid on the wall initially. It will be helpful to notice that as $h$ increases away from zero, the flux $q(h)$ is initially negative (corresponding to upward flow due to airflow), reaches a minimum of $-4/27$ when the height reaches $h=2/3$ and then increases back to zero at $h=1$ and into positive values beyond that (corresponding to downward flow due to gravity). At the same time, the wave speed $q'(h)=3h^2-2h$ also initially decreases from zero at $h=0$ to a minimum of $-1/3$ at $h=1/3$, increasing beyond that point and changing sign, becoming positive, as $h$ passes the value $h=2/3$. 

\subsection{Characteristic equations}
Define $z(s) \equiv h(x(s))$ and write the above equation along characteristics parameterized by variable $s$ as 
\begin{align}
 \frac{dt}{ds} &= 1 \\
 \frac{dx}{ds} &= 3z^2 - 2z \\
 \frac{dz}{ds} &= S(x(s))
\end{align}
If $x$ remains in the range $[0,1]$ for which $S(x) = S_0$, where $S_0 > 0$ is constant, and replacing $s$ with $t$ by assuming $s=0$ when $t=0$, we have:
\begin{align}
 z(t) &= S_0t \\ 
 x(t) &= S_0^2 t^3 - S_0 t^2 + x_0
\end{align}
where $x_0$ is the initial point along the $x$-axis where the characteristic starts, for now taken to be in the range $[0,1]$.  This solution remains valid until $x(t)$ reaches one of the boundaries $x=0$ or $x=1$. 

Starting at any value of $x_0$ in our range, the solution $x(t)$ reaches its minimal value at time  $t = {2}/{(3S_0)}$, which is independent of $x_0$. For the characteristic that starts at the bottom point $x_0=1$, this minimum would be at $x=0$ if $S_0=4/27$. Therefore, as long as 
\[
		S_0 \leq \frac{4}{27}
\]
all characteristic lines that start with $x_0 \in (0,1)$ do cross the line $x = 0$ at some finite time. Under this assumption, define $t^*$ to be the time at which a characteristic line that start within $(0,1)$ first reaches $x = 0$. Once the characteristic line crosses $x=0$, it becomes a straight line and it will not cross the $x = 0$ line again. We can calculate the straight line expression for $t > t^*$. Since we are now in the range  $x \in (-\infty , 0)$ where $S(x)=0$, the characteristic equations for $t>t^*$ become
\begin{align}
	\frac{dH}{dt} &= 0 \\
	\frac{dx}{dt} &= 3H^2 - 2H
\end{align}
where $H$ is the height function in that region, with initial conditions 
\begin{align}
	H(t^*) &= S t^* \\
	x(t^*) &= 0\,.
\end{align}
Solving these two ODEs, we have
$$
    x(t) = (3S_0^2 (t^*)^2 - 2S_0 t^*)t - 3S_0^2(t^*)^3 + 2S_0 (t^*)^2\,.
$$
At a given time $t$, we can treat the equation above as a third order polynomial with respect to $t^*$, and solve for $t^*$ for the given $t$ and $x$. Since these characteristics collide with the horizontal characteristics which emanate from the region $x \in (-\infty , 0)$, a shock forms right away at location $(x,t)=(0,0)$. If the $x$-coordinate of the shock is denoted by $c(t)$,   the Rankine-Hugoniot condition for the shock curve can be used to obtain the speed of the shock, in this case yielding:
$$
    \frac{dc}{dt} = S_0 t^* (S_0 t^* - 1)
$$
with $c(0) = 0$ and with $t^*$ a function of $t$ and $c$, obtained by solving the cubic equation given above. We applied a forward Euler method to calculate the position of the shock wave numerically. For each iteration in $t$, we solve the cubic equation to find $t^*$ and update the position of the shock.

\begin{figure}
\begin{center}
\includegraphics[scale = 0.5]{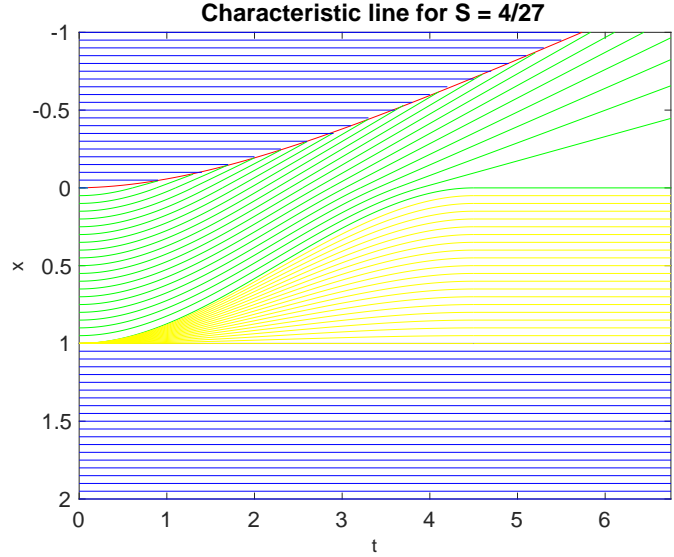}
\end{center}
\caption{A sketch of characteristic lines with $S_0 = {4}/{27}$; note that the vertical axis is the distance $x$ increasing downward, and the horizontal axis represents time $t$. The red line is the shock curve formed through the intersection of the blue and green characteristics. The blue characteristics emanate outside the source region and are horizontal. The green characteristic curves emanate from the source region and upon passing $x=0$ become straight lines. The yellow characteristics represent an expansion fan emanating from $x=1$.} 
\label{fig:characteristic_plot}
\end{figure}

Figure~\ref{fig:characteristic_plot} provides a complete picture of the characteristic curves when the source strength has its threshold value of $S_0 = \
{4}/{27}$. The numerical results show that as $t \rightarrow \infty$, the shock propagate at a constant speed of ${1}/{4}$; this is consistent with our numerical simulations of the nonlinear PDE reported below for the given source value. 

\subsection{The steady-state solution}
If a steady-state solution is reached in the region $x \in (0 , 1)$, the resulting height function must satisfy $d(h^3-h^2)/dx=S_0$, which produces the cubic equation 
\[
	h^3 - h^2 = S_0 x + C
\]
for $h(x)$. When the source strength $S_0<4/27$, the steady-state film height remains zero at $x=1$, which makes the constant $C$ equal to $-S_0$. Solving the cubic equation for $h(x)$ will then produce the correct steady-state profile over $x \in (0,1)$. 

From the method of characteristics, when the source strength exceeds the threshold, i.e., when 
\[
	S_0 \geq \frac{4}{27}\,,
\]
the characteristic emanating from the initial point  $x_0=4/(27S_0) \in (0,1)$ becomes tangent to the horizontal line $x=0$ at time $t=2/(3S_0)$, at which point $h(0, {2}/{(3S_0)}) = {2}/{3}$. Beyond that time, the height at that location remains constant at value 2/3,  which enables us to determine the constant $C$ for that case. Also in that case, the characteristics starting at $x_0>4/(27S_0)$ (but less than 1), do not reach $x=0$ at any time and instead turn around and exist the domain at $x=1$, colliding with the horizontal characteristics that emanate from the region $x_0>1$. This leads to a second shock front that propagates downward, reflecting the fact that at high source values, some of the flow is carried downward by gravity.

\begin{figure}[h]
\begin{center}
\includegraphics[scale = 0.5]{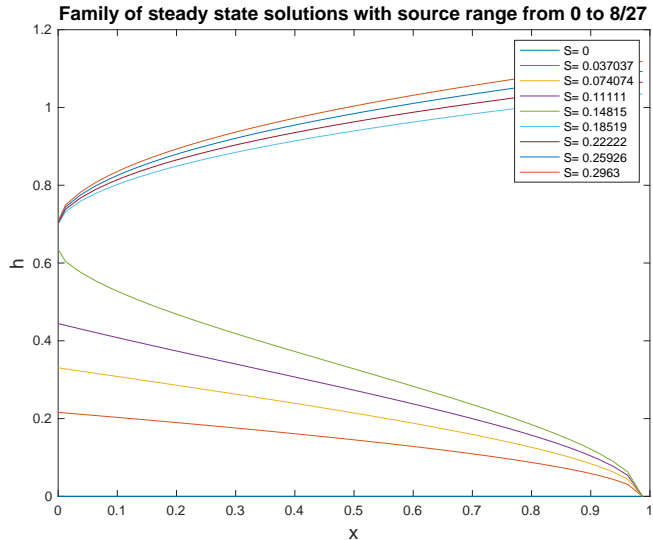}
\end{center}
\caption{The family of steady state solutions with $S_0 $ ranging from 0 to ${8}/{27}$. The bottom curves are for source strengths below the threshold and the top curves for those above the threshold. } 
\label{fig:family_of_steady_state_sol}
\end{figure}

In order for the steady height to remain constant equal to 2/3 at $x=0$, the constant $C$ must be given by: 
\[
 	C = 	(\frac{2}{3})^3 - (\frac{2}{3})^2 = -\frac{4}{27}\,.
\]
Then steady-state height profile for $S_0 \geq {4}/{27}$ would be the solution of the new cubic equation: 
\[
	h^3 - h^2 = S_0 x - \frac{4}{27}\,.
\] 
The two cases for $S_0$ below and above the threshold can be combined to write a single cubic equation whose solution provides the steady-state film profile $h(x)$: 
\begin{equation}
	h^3 - h^2 = S_0 x - \min \{ S_0, \frac{4}{27} \} \,.\label{steady_state}
\end{equation}
For source strengths below the threshold 4/27, the steady height remains constant equal to zero at x=1, and for those above the threshold, the steady height remains constant equal to 2/3 at x=0. These can be verified from the numerical simulation of the nonlinear PDE which is described next. Figure~\ref{fig:family_of_steady_state_sol} provides a plot of the family of steady state film profiles over $x \in (0,1)$ for source values below and above the threshold $\frac{4}{27}$.

\subsection{Numerical simulations}
For our simplified model without surface tension, we now describe the Godunov method that provides a numerical solution for the time evolution of the film thickness. 

\subsubsection{Godunov method}
We discretize the $x$-domain into $N$ equally-spaced sub-intervals or cells of size $\Delta x$ with point $x_j$ referring to the midpoint of the cell $j$, whose edges are at $x_{j-\frac{1}{2}}=x_j - \Delta x/2$ and $x_{j+\frac{1}{2}}=x_j + \Delta x/2$. Time domain $t$ is also discretized with time-step $\Delta t$ so that $t_n=n \Delta t$. We denote the average film thickness over cell $j$ at time level $n$ by
\[
	H_j^n = \frac{1}{ \Delta x} \int_{x_{j-\frac{1}{2}}}^{x_{j+\frac{1}{2}}} h(x, t_n) dx \,.
\]

We integrate the conservation equation $h_t+q_x=S(x)$ (with $q=q(h)=h^3-h^2$) over the domain $[x_{j-\frac{1}{2}}, x_{j+ \frac{1}{2}}] \times [t_n, t_{n+1}]$ and simplify to obtain
\[
	H_j^{n+1} = H_j^{n} - \frac{1}{\Delta x} \int_{t_n}^{t_{n+1}} q\big(h(x_{j + \frac{1}{2}},t)\big) - q\big(h(x_{j + \frac{1}{2}},t)\big) dt + \frac{\Delta t}{ \Delta x} \int_{x_{j-\frac{1}{2}}}^{x_{j+\frac{1}{2}}} S(x) dx\,.
\]
Denote the time-average of the flux crossing the edge $x_{j+\frac{1}{2}}$ over the time interval $t \in [t_n,t_{n+1}]$ as 
\[
	\overline{Q}_{j + \frac{1}{2}}^n = \frac{1}{\Delta t} \int_{t_n}^{t_{n+1}} q(u(x_{j + \frac{1}{2}},t)) dt \,,
\]
which produces the discrete conservation equation
\[
	H_j^{n+1} = H_j^{n} - \frac{\Delta t }{\Delta x} \left( \overline{Q}_{j + \frac{1}{2}}^n - \overline{Q}_{j - \frac{1}{2}}^n + \int_{x_{j-\frac{1}{2}}}^{x_{j-\frac{1}{2}}} S(x) dx \right)\,.
\]

In Godunov's method, the time-averaged flux $\overline{Q}^n_{j+\frac{1}{2}}$ is approximated as follows 
\[
	\overline{Q}^n_{j+ \frac{1}{2}} = Q(H_{j}^n, H_{j+1}^{n}) = \begin{cases}
		\min_{H_{j}^n \leq \theta \leq H_{j+1}^{n} } q(\theta) &\text{if} \quad H_{j}^n \leq H_{j+1}^{n}\\
		\max_{H_{j+1}^n \leq \theta \leq H_{j}^{n} } q(\theta) &\text{if} \quad H_{j}^n > H_{j+1}^{n}\\
	\end{cases}
\]
relating the flux to the average heights on either side of the edge at time level $n$. In our case, since $q(h) = h^3 - h^2$ and $h \geq 0$, the only minimum in $q(h)$ occurs at $h=2/3$ and the formula simplifies to
\[
	\overline{Q}^n_{j+ \frac{1}{2}} = \max \Big( q\big( \max (H_j^n , \frac{2}{3}) \big), q\big( \min (H_{j+1}^n , \frac{2}{3})\big)\Big)\,.
\]
For numerical stability, one must require the time step to be small enough, according to the stability condition 
\[
	\frac{\Delta t}{\Delta x} \leq \frac{1}{2 \max_j |q'( H_j^n )| }\,.
\]
In the simulations presented below, we take ${\Delta t}/{\Delta x} = {1}/{8}$.

\subsubsection{Results}
In the following, we present results for the case $S_0=5/27$, which is above the threshold value of $4/27$. We thus expect some of the flow to be carried downward by gravity, while some portion is still carried upward by the airflow. We simulate the equation over the region $x\in[-5,5]$ with $\Delta x = 0.025$. 

\begin{figure}
\begin{center}
\includegraphics[scale=0.5]{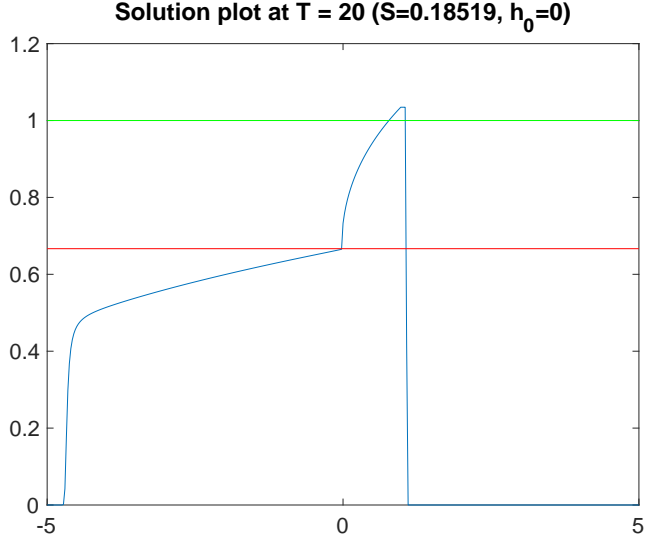}
\end{center}
\caption{Plot of the numerical solution at time $T = 20$, with $S_0 = \frac{5}{27}$ and $h_0=0$. The horizontal axis represents the $x$-coordinate along the vertical wall, with the positive direction being downward. The vertical axis is the height of fluid film. Since this source value is larger than critical value $\frac{4}{27}$, we can see two shock waves, one going upward and one downward.}
\label{fig:final_step}
\end{figure}

Figure~\ref{fig:final_step} presents the film profile at time $t=20$ starting with no liquid film for a source strength of $S_0=5/27$ acting over $x\in[0,1]$. The horizontal lines at heights 1 and 2/3 are drawn for visual references. Once the film height reaches a value of 2/3 at $x=0$, it stays at that value, while the excess liquid is carried upward (toward negative $x$ values) by the airflow. Some of the liquid also flows downward (toward positive $x$ values) due to gravity although at time $t=20$, only a small amount has gone past the edge $x=1$. 

\begin{figure}
\begin{center}
\includegraphics[scale=0.5]{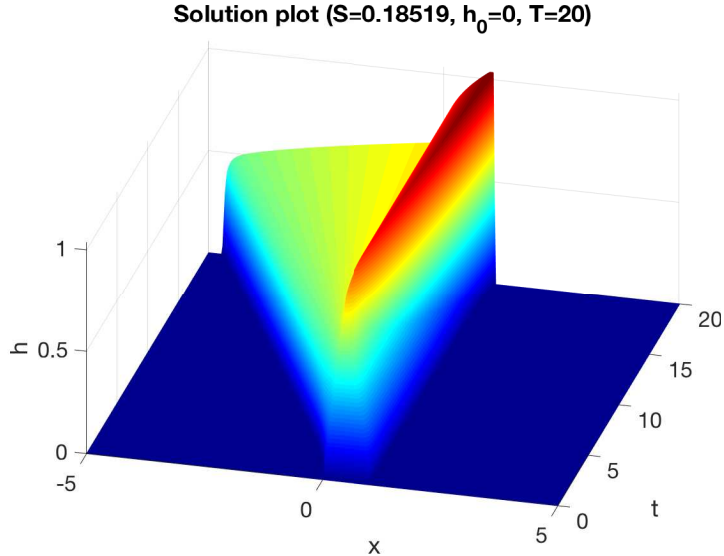}
\end{center}
\caption{Evolution of the film profiles from $T = 0$ to $20$ for $S_0 = \frac{5}{27}$ and $h_0 = 0$. The film height is plotted as a function of $x$ and $t$.}
\label{fig:evolution}
\end{figure}

Figure~\ref{fig:evolution} presents the evolution of the film profile from time $t=0$ to $t=20$ starting with zero initial film thickness and with a source strength of $S_0=5/27$ confined to the region $x \in [0,1]$. It is seen that the shock traveling upward (toward negative $x$ values) achieves a fairly constant speed of propagation.

\begin{figure}
\begin{center}
\includegraphics[scale=0.5]{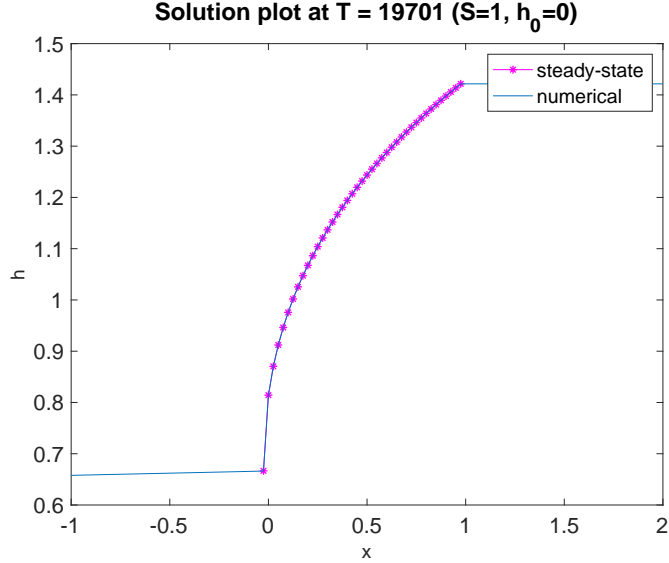}
\end{center}
\caption{Comparison of numerical simulation results at large times with the steady state solution calculated from Eq.~(\ref{steady_state}) over the range $x \in [0,1]$.}
\label{fig:steady_state_comparison}
\end{figure}

In Figure~\ref{fig:steady_state_comparison} we compare the numerical solution at large times over the range $x \in [0, 1]$ to the steady state solution over that range which solves Eq.~(\ref{steady_state}). The two results are in excellent agreement. 

\section{Model with Surface Tension}
In Section~\ref{sec:derive} we derived the model with surface tension in the form of Eq.~(\ref{PDE_with_alpha}) in which dimensionless parameter $\alpha=\sigma H /(\rho g L^3)$ measured the relative importance of surface tension. While the previous section analysed the system when $\alpha=0$, here we will examine the solution when that parameter is nonzero.


\subsection{Numerical Simulations using COMSOL}

For the full model with surface tension, we conduct numerical simulations using the software COMSOL MultiPhysics for various source strengths $S_0$. We observe some similarities with the simplified model; however, there are significant differences also. 


\begin{figure}
    \centering
    \includegraphics[width=0.9\textwidth]{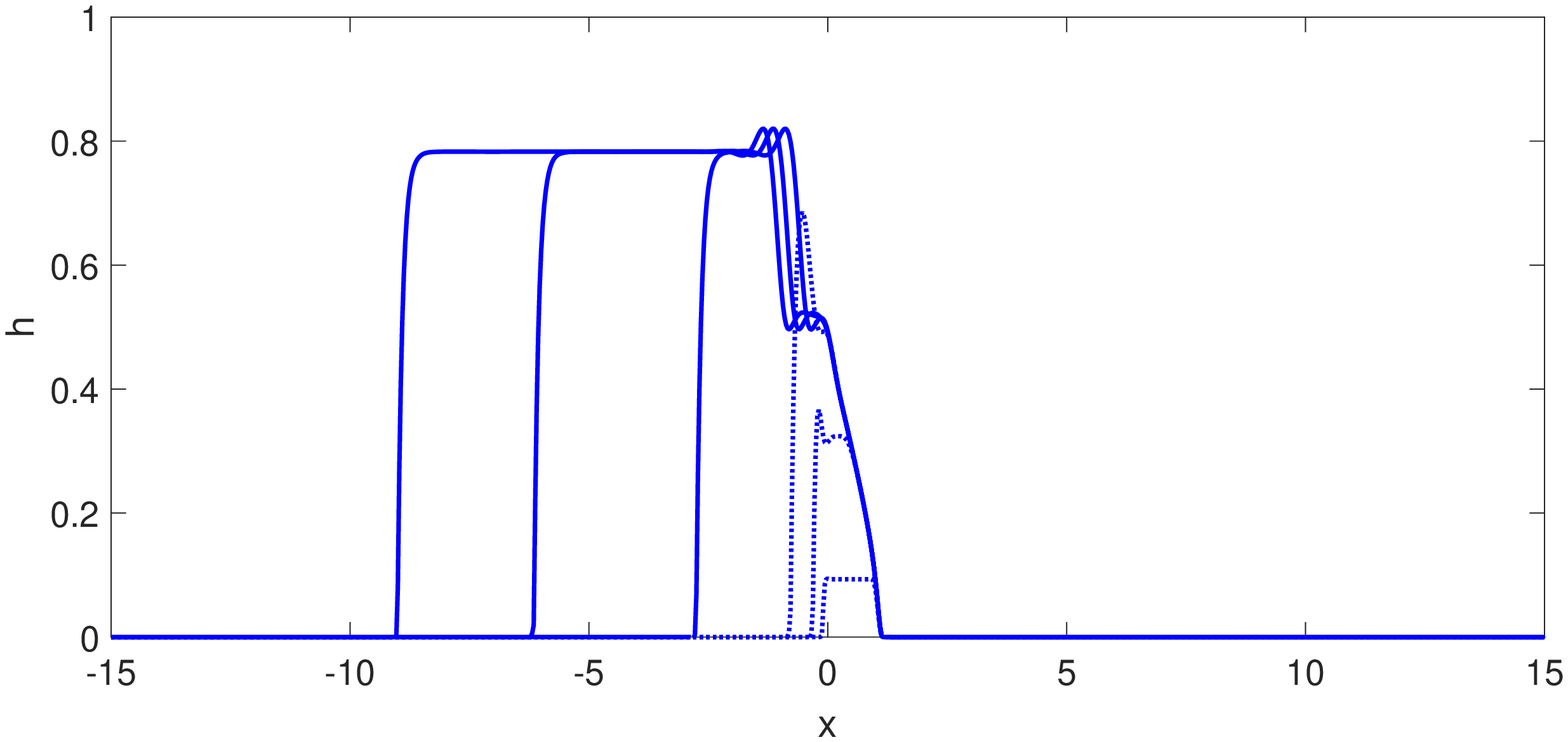}
     \includegraphics[width=0.9\textwidth]{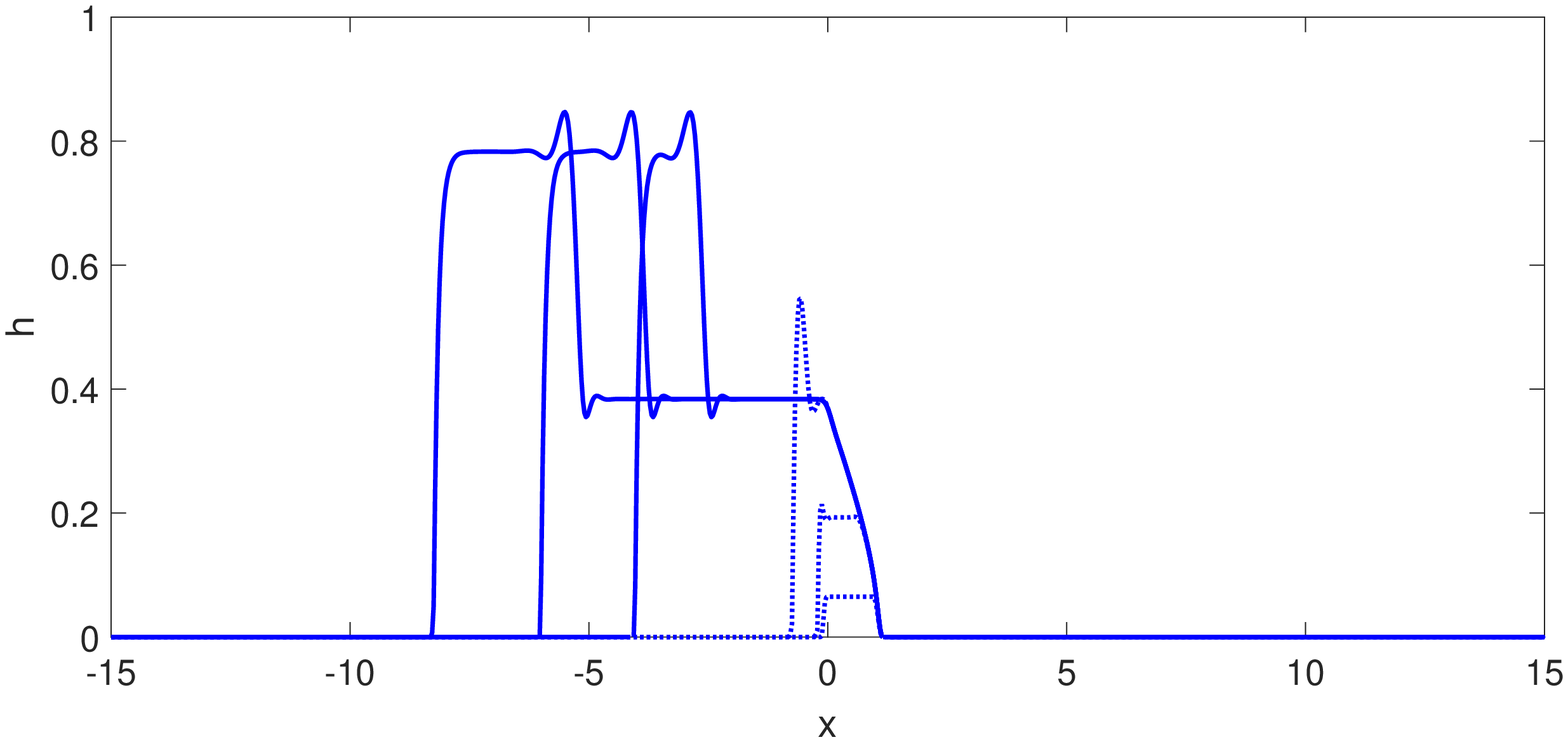}
    \includegraphics[width=0.9\textwidth]{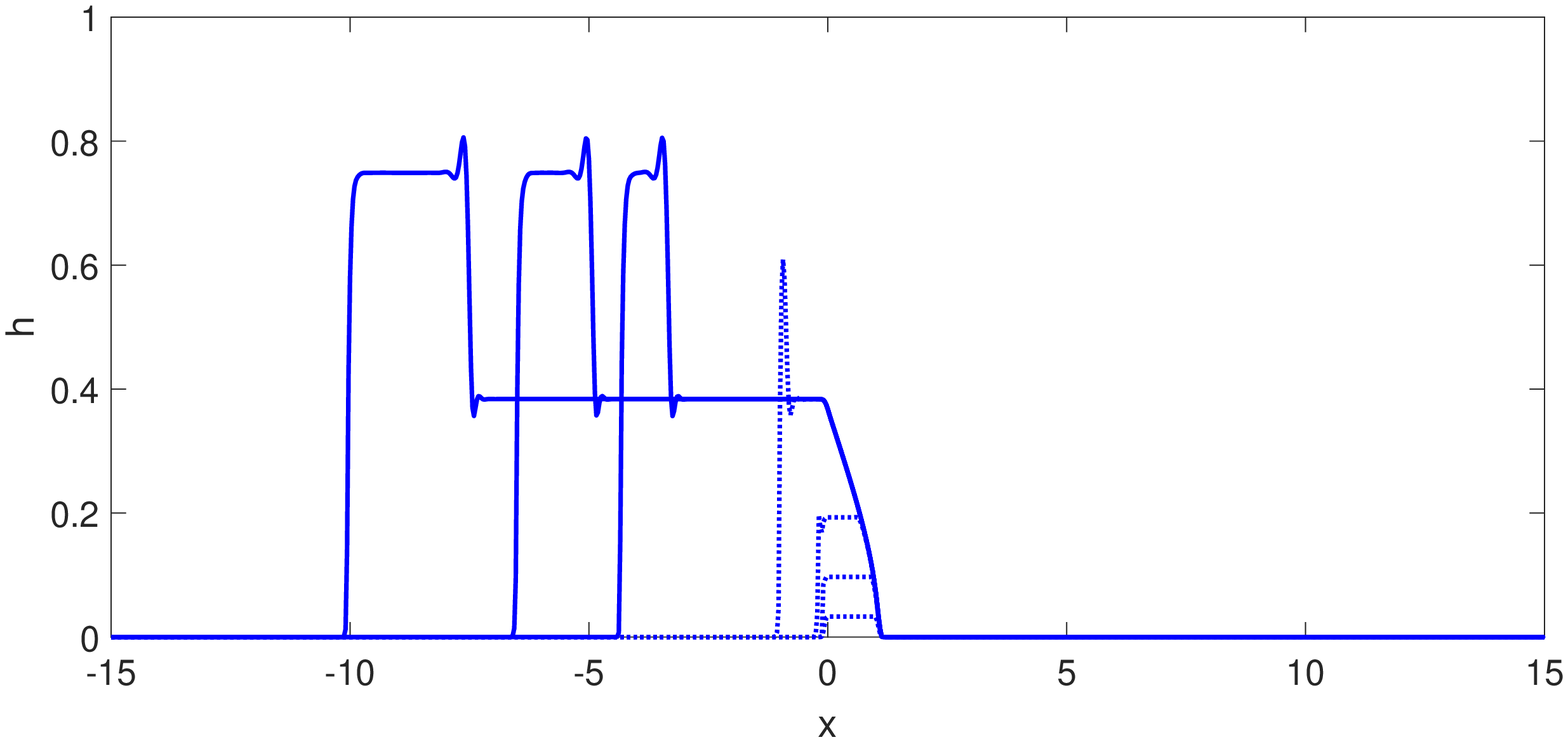}
    \caption{Weak source simulations indicate the propagation of waves in one direction only: to the left due to airflow. Top picture: $S_0=4/35$, $\alpha = 0.001$; middle picture:  $S_0=4/50$, $\alpha = 0.001$; bottom picture: $S_0=4/50$, $\alpha = 0.0001$. Refer to the text for more detailed descriptions.}
    \label{fig:weak}
\end{figure}

Figure~\ref{fig:weak} provides a series of simulations over the domain $(-15,15)$ corresponding to weak source strengths (below the threshold of $4/27$ predicted for zero surface tension). In the top picture, the dashed lines present the early time evolution snapshots for parameter values: $S_0=4/35$ and $\alpha = 0.001$, at times: $t = 0.8; 2.8; 5.6$. The solid lines present the later time evolution snapshots at times $t = 17.2; 37.2; 54.0$. A steady state is established over the source region $(0,1)$. For these weak source values, none of the fluid falls due to gravity (i.e., none moves to the right beyond the edge $x=1$). The fluid that gets transported to the left (upward due to airflow) has a leftmost front that looks like a typical shock, but the relatively flat region next to that front jumps down to a lower value (more evident in the middle and bottom panels) across an oscillatory front that propagates at a different speed from the leftmost front. The left front wave has height $0.783$, moving to the left at speed $0.166$; the second left wave is very slow with a speed of about $0.012$ and a peak height of about $0.845$. The height of the left front wave does not depend on the source strength, as will be seen in the middle panel.

In the middle picture, the dashed lines provide the early time evolution snapshots for $S_0=4/50$ and $\alpha = 0.001$ at times: $t = 0.8; 2.4; 6$. The solid lines show the later time evolution snapshots at times: $t = 25.2; 36.8; 50.0$. The left front wave has height $0.783$ and moves to the left with speed $0.173$. The second left-going wave is slower with a speed of about $0.107$ and a peak height of about $0.845$, connecting to a flat part of height $0.384$. Changing the source strength does not influence the height of the left front wave but it moves a bit faster, the height of the second left wave is also unchanged but it is moving much faster to the left as we decrease the source strength away from its threshold value.

In the bottom panel, the dashed lines give the early time evolution snapshots for an even smaller surface tension case, with parameter values: $S_0=4/50$ and $\alpha = 0.0001$ at times: $t = 0.4; 1.2; 2.4; 7.2$. The solid lines provide the later time evolution snapshots at times $t = 24.8; 36.4; 55.2$. The left front wave has height $0.749$ and moves with speed $0.188$. The second left wave is slower with a speed of about $0.138$ and a peak height of about $0.806$, connecting to a flat part of height $0.384$. Reducing surface tension speeds up the front left wave and lowers its height, but contrary to the strong source case (presented next) the second wave also speeds up.

It follows from the simulations that the height of the left front wave is not controlled by the source term and only depends on the surface tension coefficient, as does its speed. The second wave speed and direction, however, are controlled by the source strength.

\begin{figure}
    \centering
    \includegraphics[width=0.9\textwidth]{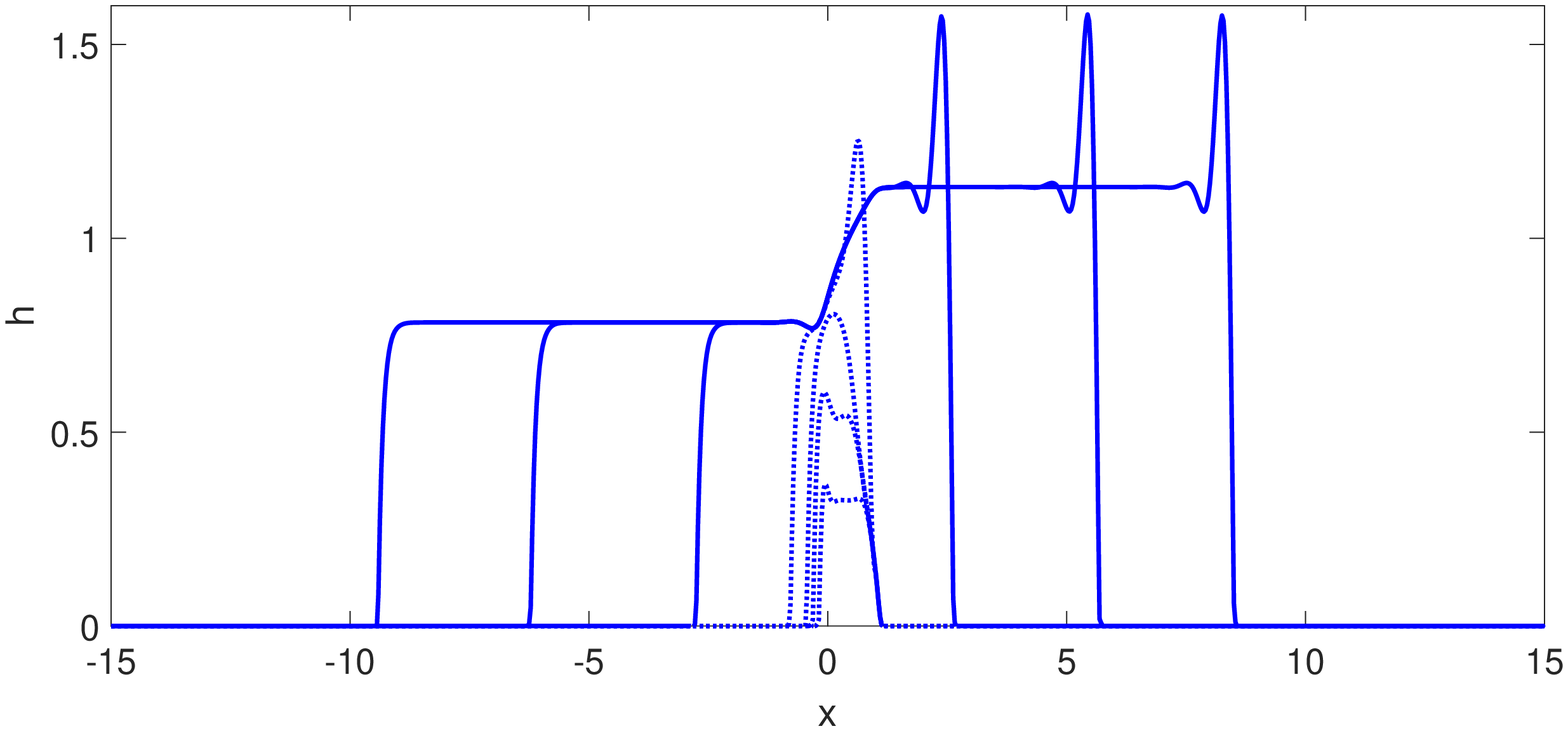}
    \includegraphics[width=0.9\textwidth]{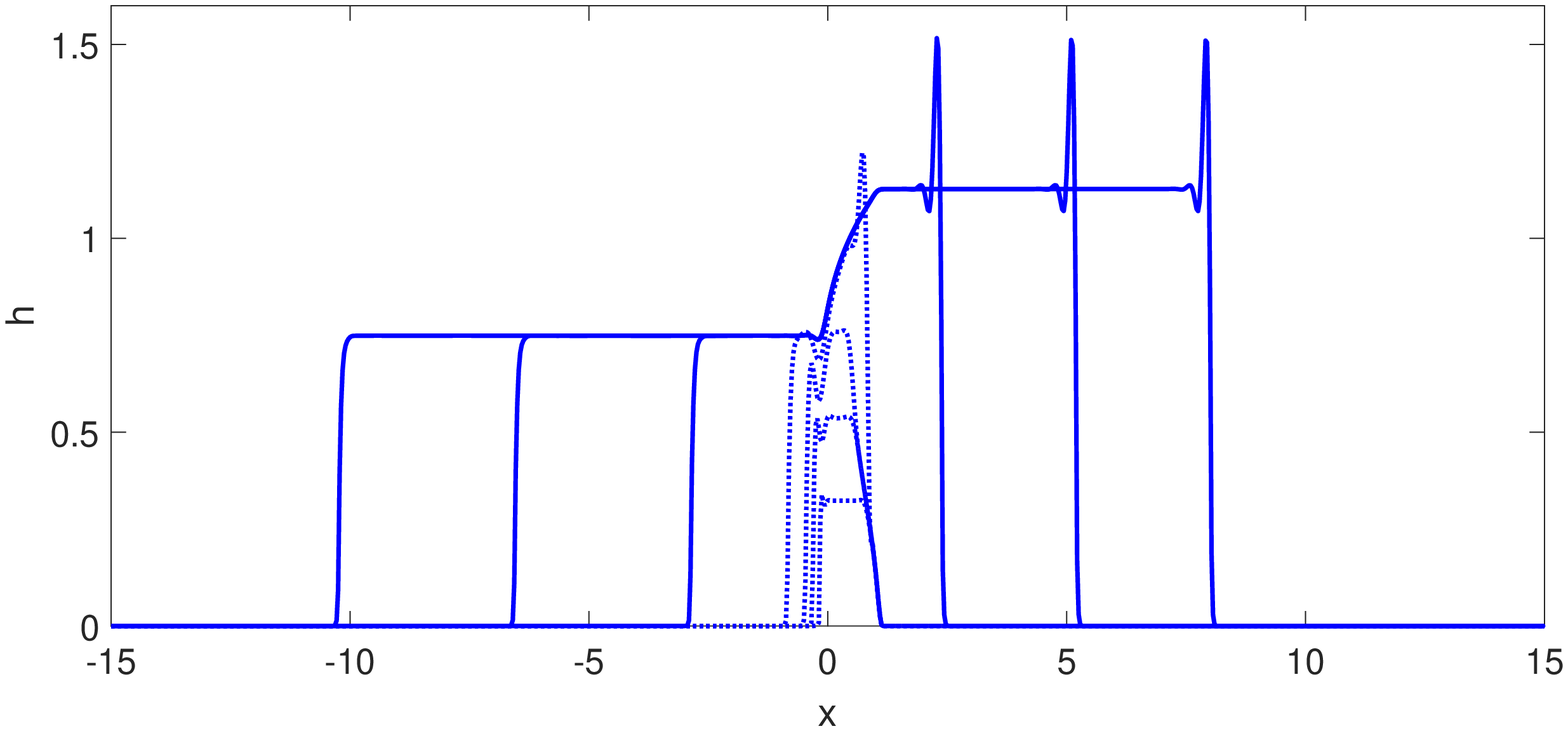}
     \includegraphics[width=0.9\textwidth]{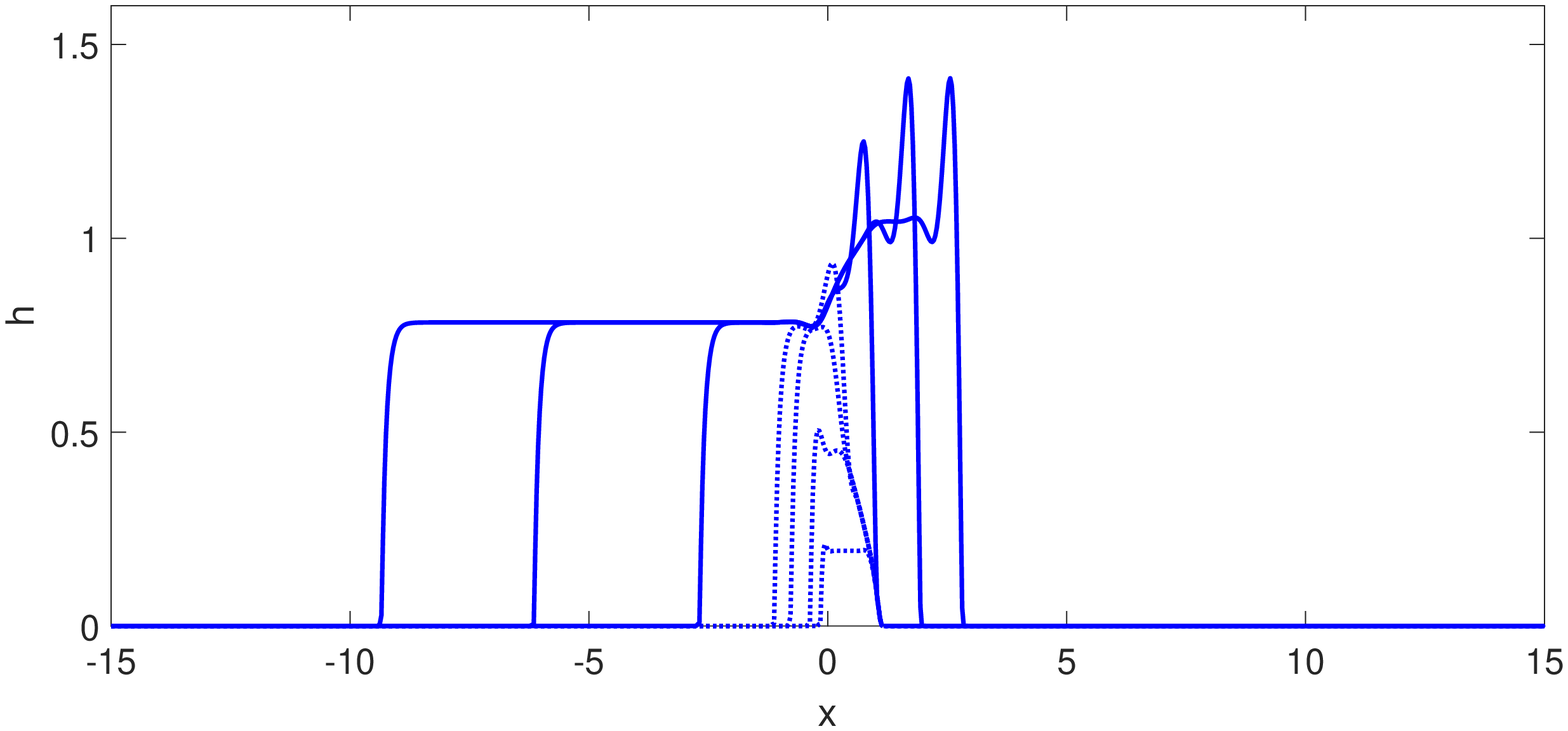}
    \caption{Strong source simulations indicate propagation of waves in both directions: left-going due to airflow and right-going due to gravity. Top picture: $S_0=4/15$, $\alpha = 0.001$; middle picture: $S_0=4/15$, $\alpha = 0.0001$; bottom picture: $S_0=4/25$, $\alpha = 0.001$. Refer to the text for more detailed descriptions.}
    \label{fig:strong}
\end{figure}

Figure~\ref{fig:strong} presents a set of simulations with stronger source strengths (above the threshold) that result in two fronts moving in opposite directions. In the top panel, the dashed lines provides the early time snapshots for parameter values: $S_0=4/15$, $\alpha = 0.001$, at times $t = 1.2; 2.0; 2.8; 4.8$. The solid lines are the later time snapshots at $t = 16.4; 36.8; 55.6$. The left-going front has a height of approximately $0.783$, moving with an approximate speed of $0.168$. The right-going front has a flat part of height $1.132$ (peak at $1.573$) and it is moving a little slower with a speed of about $0.163$. 

In the middle panel, the dashed lines are the early time snapshots for parameter values:
$S_0=4/15$ and $\alpha = 0.0001$ (a factor of ten smaller than the top panel) at $t = 1.2; 2.8; 5.2; 7.2$. The solid lines are the later snapshots for parameter at times $t = 15.6; 35.2; 54.8$. The left-going front has an approximate height of $0.749$ and moves with an approximate speed of $0.188$. The right-going front has a flat region of height $1.127$ (peak at $1.510$) and it is moving a little slower at a speed of about $0.143$. The ten-fold reduction in surface tension from $0.001$ to $0.0001$ resulted in a faster propagation of the left front while lowering its height. The right-moving front also has a lower height but, contrary to the left wave, it slows down.

In the bottom panel, the dashed lines are the early time snapshots for parameter values:
$S_0=4/25$ and $\alpha = 0.001$, at $t = 1.2; 2.8; 5.2; 7.2$. The solid lines are the later time snapshots at times $t= 16.4; 36.8; 55.6$. The left-going front has height $0.783$ and moves with speed $0.173$. The right-going front has a flat part of height $1.132$ (peak at $1.573$) and it moves slower at a speed of about $0.047$.  Reducing the source from $4/15$ to $4/25$ does not affect the height of the left-going front but increases its speed, while the right-moving front does not change its height either, but slows down appreciably. 

By examining both Figures~\ref{fig:weak} and \ref{fig:strong} combined, it becomes apparent that for weak source strengths below the threshold, we have the left-going front and the second left-moving wave. The latter moves to the left more and more slowly as the source strength approaches the threshold and eventually changes directions and becomes a right-moving wave as the source-strength increases above the threshold value. While both of these waves exhibit oscillations before connecting two flat regions, when the wave moves to the right (due to gravity), it connects a flat region of zero, whereas when it was moving to the left, it connected two flat regions of finite heights.

Another important observation one can make by comparing the results with surface tension with those in the complete absence of surface tension (i.e., Figure~\ref{fig:final_step} from the previous section), is that even a small amount surface tension ($\alpha=0.001$) appreciably slows down the left-moving front and makes the region behind the left front flat, as opposed to having a clear slope apparent in Figure~\ref{fig:final_step}.

\begin{figure}
    \centering
    \includegraphics[width=1.0\textwidth]{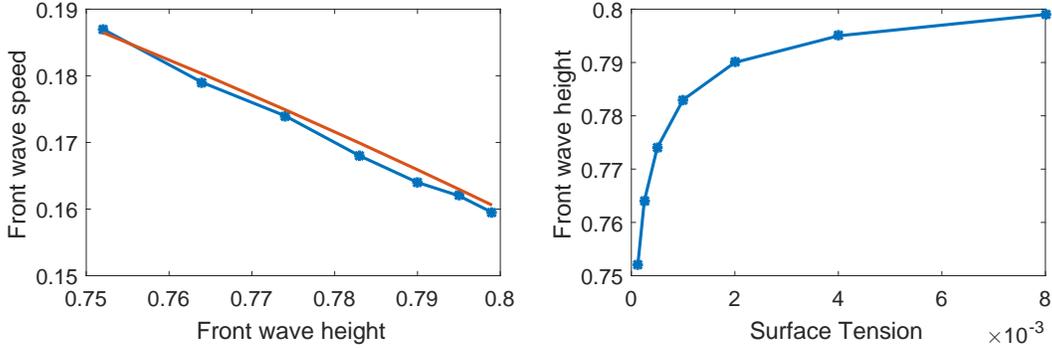}
    \caption{The plot of the left front wave speed versus its height is on the left and the plot of the left front wave height versus values of $\alpha$ (surface tension coefficient) is on the right. }
    \label{fig:plot2}
\end{figure}

To explore the effect of surface tension on the front propagation, we examine the height and speed of the left-going wave for a larger set of surface tension parameter values $\alpha$. Figure~\ref{fig:plot2} ($S_0 = 4/15$) shows that left-going front speed decreases with height 
and that as surface tension parameter $\alpha$ becomes larger, the front height approaches an approximate value of $0.8$ (for an even higher value of $\alpha = 0.1$ the height is about $0.806$, and for $\alpha = 0.5$ the height is about $0.805$). 
The relation between the front speed and height seen on the left plot in Fig.~\ref{fig:plot2} can be explained by seeking a traveling wave solution of Eq.~(\ref{PDE_with_alpha}) away from the source region. For a left-going wave, if we take $h(x,t)$ to have the traveling wave form $h(x+ct)$ with $c>0$, the PDE away from the source reduces to
\[ c h + (h^3-h^2+\alpha h^3 h''')=\mbox{constant} \,. \]
In the flat regions on either side of the front, $h'''$ is zero, and to the left of the front, $h=0$. This makes the constant on the RHS equal to zero and for the flat region of height $h$, the traveling wave speed is evaluated to be $c=h-h^2$ (see the red curve on the figure). This is in approximate agreement with the data points plotted in the figure. 

\section{Discussion and Conclusions}
\begin{figure}
    \centering
    \includegraphics[width=0.9\textwidth]{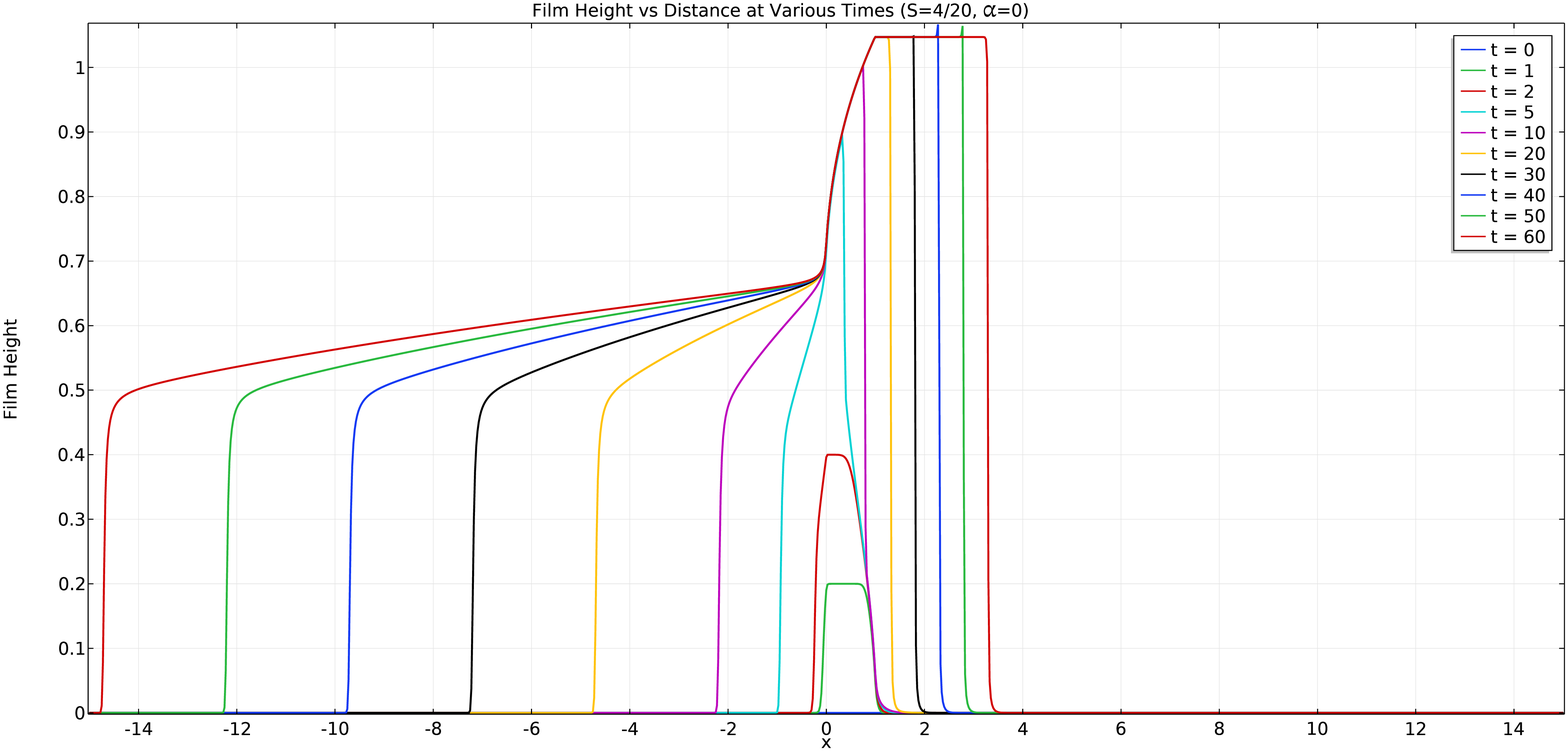}
    \includegraphics[width=0.9\textwidth]{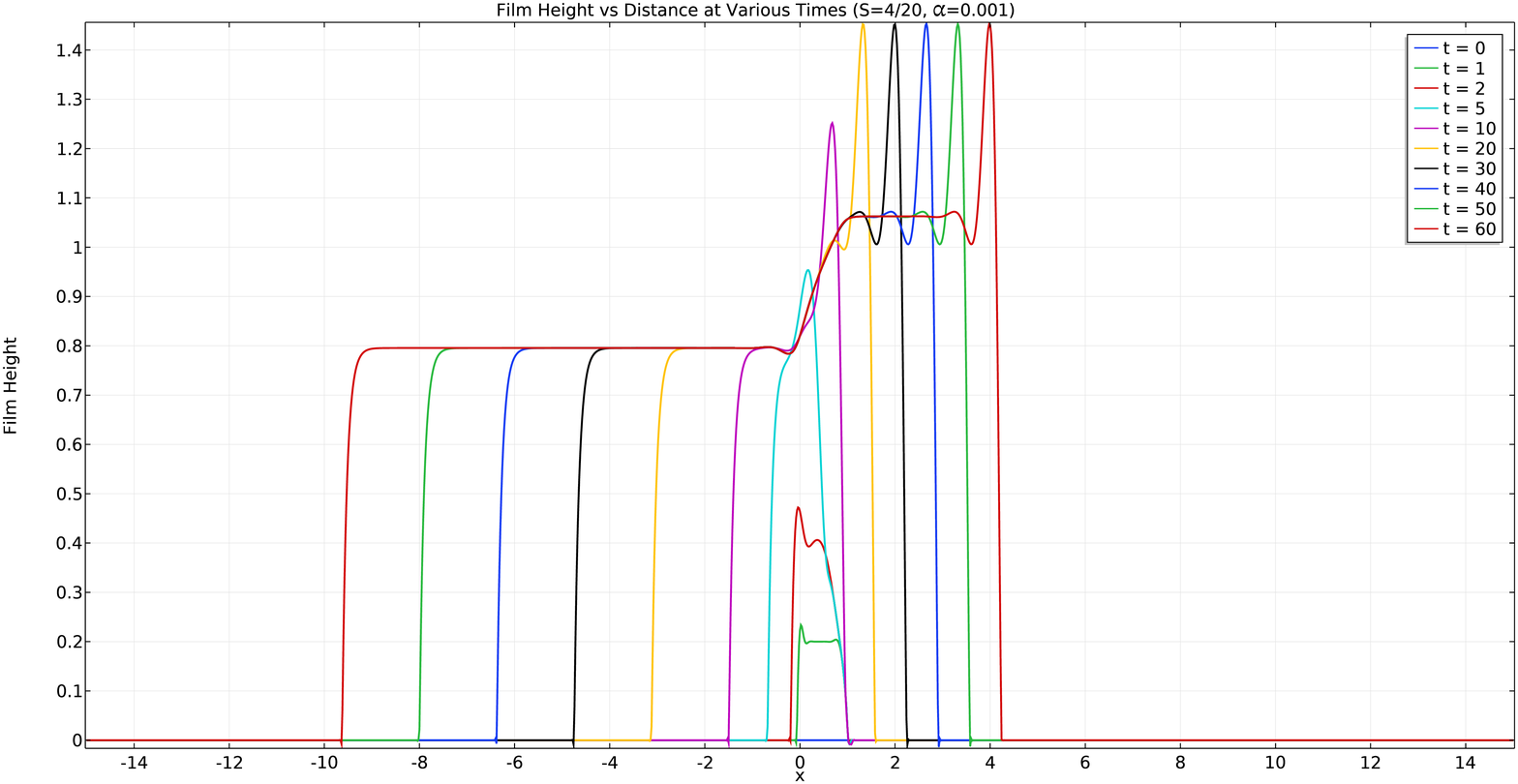}
    \caption{Comparison of models without and with surface tension.}
    \label{fig:comparison}
\end{figure}

Let us first compare the cases with and without surface tension to highlight their key differences. Figure~\ref{fig:comparison} shows two sets of simulations for a source strength of $S_0=4/20$ that produces traveling waves in both direction, with liquid climbing the wall due to airflow (going left in the plots) and excess liquid falling down due to gravity (going right in the plots). In the top panel, surface tension is zero ($\alpha=0$), while in the bottom panel surface tension is nonzero but rather small ($\alpha=0.001$). The profiles are plotted at the same times indicated in the legend. It is obvious that even for quite small values of the surface tension parameter, the profiles are strongly affected. In the absence of surface tension, the left-going waves in the top figure advance at a higher speed and the profiles behind them have positive slopes that decrease as the front advances. In contrast, in the presence of surface tension, that front moves left more slowly and behind the front, the profile is flat and maintains a constant somewhat higher height. On the other hand, the right going waves move a little faster when surface tension is present, and the constant part of the profile behind those waves connects to the zero region in front through an oscillatory section with a large peak, to be compared to the flat profile of the right-going waves without surface tension in the top figure. Since the surface tension parameter $\alpha$ multiplies the highest (fourth order) spatial derivative term in the governing equation, it is not too surprising that from a perturbation standpoint, the problem is singular and even quite small values of the surface tension parameter $\alpha$ significantly modify the behavior of the solution. 

When the source strength is large enough, this model generates two travelling waves moving left and right away from the source region, connected through a steady state film profile directly over the source area. For sub-threshold source strengths, only left-going waves are observed, but there are two such waves that travel at different speeds. For any of the traveling waves that connect two flat regions (one possibly of zero height far to the left or right), a Rankine-Hugoniot equation can be obtained that relates the speed of the moving front to the constant heights on either side of the traveling ``shock.'' This is easy to see by substituting a travelling wave ansatz $h(x,t) = h(z), z = x - c t$ into the PDE 
$$
	h_t + (h^3 - h^2 + \alpha h^3 h_{xxx})_x = 0
$$
away from the source region. This yields
$$
	-c h' + (h^3 - h^2 + \alpha h^3 h''')' = 0
$$ with the prime denoting a $z$-derivative. Integrating the equation once, we have
$$
	h^3 - h^2 + \alpha h^3 h''' = c h + C\,,
$$
where $C$ is an integration constant. When a travelling wave connects uniform left and right regions with heights $h_{-}$ and  $h_{+}$, since $h(z)''' = 0$ as $z \rightarrow \pm \infty$, we find that
$$
	h_{-}^3 - h_{-}^2 - c h_{-} = h_{+}^3 - h_{+}^2 - c h_{+} = C\,.
$$
The wave speed $c$ can thus be obtained: 
$$
	c = \frac{ (h_{-}^3 - h_{-}^2) - (h_{+}^3 - h_{+}^2)}{ h_{-} - h_{+}} = h_{-}^2 + h_{-}h_{+} + h_{+}^2 - h_{-} - h_{+}
$$
This is consistent with the Rankine-Hugoniot condition that $c=[\![q(h)]\!]/[\![h]\!]$, where $q(h)$ is the flux and double square brackets indicate the jump in the value of their argument from one side to the other. 

From our numerical simulation results, we can verify that the travelling wave speed is indeed given by this equation. For instance, consider the small source condition depicted in the middle panel of Figure~\ref{fig:weak}. We see two travelling waves both traveling to the left. Denote the flat part of the height profile from left to right as $h_1, h_2, h_3$; in that case:
$$
	h_1 = 0,\quad h_2 = 0.783, \quad h_3 = 0.384\,.
$$
Denote the two wave speed from left to right as $c_1$ and $c_2$. The predicted wave speeds would thus be:
\begin{align*}
	c_1 &= h_{1}^2 + h_{1}h_{2} + h_{2}^2 - h_{1} - h_{2} = -0.170 \\
	c_2 &= h_{2}^2 + h_{2}h_{3} + h_{3}^2 - h_{2} - h_{3} = -0.106\,.
\end{align*}
These values closely match the results obtained from studying the plot and extracting the velocities. 



In order to get some sense of the orders of magnitude of the parameters and the applicability of the lubrication approximation, let us consider a hypothetical case with the following physical parameters. Take the liquid and gas to be water and air at 25$^{\circ}$C with respective properties: $\rho_w=997$~kg/m$^3$, $\mu_w=8.9\times 10^{-4}$~kg/(m~s), $\rho_a=1.18$~kg/m$^3$ and $\mu_a=1.85\times 10^{-5}$~kg/(m~s). Take the upward airflow velocity to be $U_a=15$~m/s and suppose that a uniform flow of that speed encounters the vertical plate, developing a laminar (Blasius) boundary layer, reaching the liquid film about $\ell=5$~cm from the bottom of the plate. In that case, the wall shear stress is given from the standard expression for a laminar boundary layer on a flat plate, namely: 
\[ \tau = 0.332 \rho_a U_a^2 / \sqrt{Re_\ell} \]
in which the Reynolds number for the airflow is defined by $Re_\ell=\rho_a U_a \ell/\mu_a$. The resulting shear stress turns out to be $\tau=0.404$~kg/(m~s$^2$). The characteristic thickness of the film which is determined by a balance of gravity and airflow is thus calculated to be
\[ H = \frac{3 \tau}{2 \rho_w g} \approx 62~\mbox{microns}\]
and if the length of the source region is taken to be $L=1$~cm, the lubrication parameter will be $\epsilon=H/L\approx 0.0062 \ll 1$. The velocity scale for the downward draining of the water film under gravity is given by $U = \rho_w g H^2/\mu_w \approx 0.042$~m/s, making the Reynolds number for water flow to be $Re = \rho_w U L/\mu_w \approx 472$. While this value is not small, the product $\epsilon^2 Re = 0.018 \ll 1$, so the neglect of inertial terms in the thin film equation can be justified. Based on these values, the dimensional threshold value for the source strength is found to be
\[ S_0 = \left(\frac{4}{27}\right)\left(\frac{9 \tau^3}{8 \mu_w \rho_w^2 g^2 L}\right) \approx 12.9~\mbox{microns per second}.\]
Finally, for the dimensionless parameter $\alpha=\sigma H/(\rho_w g L^3)$ to have value $10^{-4}$, the surface tension would have to be $\sigma=0.0158$~kg/s$^2$, or ten times higher if $\alpha=10^{-3}$. This is in the right range for water which has a surface tension of about 0.072~kg/s$^2$. So, although the dimensionless surface tension parameter $\alpha$ is indeed small for water, our analysis shows that the thin films that advance upward due to airflow, or fall due to gravity are still significantly affected by surface tension.

\textbf{Acknowledgment}: The authors would like to thank the organizers and participants of the 2018 Mathematical Problems in Industry Workshop (Claremont, California, 2018) where the problem studied in this article was originally proposed. 


%

\end{document}